\documentclass[journal, twoside]{IEEEtran}

\usepackage{cite}
\usepackage[pdftex]{graphicx}
\usepackage[hyphens]{url}
\usepackage{subfig}
\usepackage{soul}

\begin{document}

\title{Security and Privacy Analyses of\\Internet of Things Children's Toys}

\author{Gordon~Chu,
        Noah~Apthorpe,
        and~Nick~Feamster%
\thanks{Manuscript received May 9, 2018; revised June 20, 2018.}%
\thanks{G.~Chu, N.~Apthorpe, and N.~Feamster are with the Computer Science Department, Princeton University, Princeton,
NJ, 08540 USA (e-mail: gordon.chu@princeton.edu, apthorpe@cs.princeton.edu, feamster@cs.princeton.edu).}%
}


%




\maketitle

\begin{abstract}
This paper investigates the security and privacy of Internet-connected children's smart toys through case studies of three commercially-available products. We conduct network and application vulnerability analyses of each toy using static and dynamic analysis techniques, including application binary decompilation and network monitoring. We discover several publicly undisclosed vulnerabilities that violate the Children's Online Privacy Protection Rule (COPPA) as well as the toys' individual privacy policies. These vulnerabilities, especially security flaws in network communications with first-party servers, are indicative of a disconnect between many IoT toy developers and security and privacy best practices despite increased attention to Internet-connected toy hacking risks. 
\end{abstract}


\IEEEpeerreviewmaketitle

\section{Introduction}
\IEEEPARstart{I}{nternet}-connected children's toys are a subset of IoT devices that merit special attention from the security community. 
If IoT toys are compromised, a cyber predator could communicate with or collect sensitive data about children without being physically near them.
The 2015 Hello Barbie hack, in which security researchers were able to remotely access the doll's microphone and record children's conversations \cite{guardian:barbie}, clearly demonstrates the importance of securing IoT toys. 

Regulators have given special attention to privacy concerns of online services targeted toward children. The Federal Trade Commission's Children's Online Privacy Protection Rule (COPPA) places specific requirements on these services, including that they must ``establish and maintain reasonable procedures to protect the confidentiality, security, and integrity of personal information collected from children'' \cite{coppa}.

In addition to mandated COPPA compliance, manufacturers of IoT toys provide their own privacy policies to indicate data handling practices and security measures specific to their products. 
However, consumers have no way to verify whether IoT toys actually follow COPPA regulations or manufacturer privacy policies. Recent high-profile hacks of IoT devices give reason to doubt the security claims of these products~\cite{guardian:barbie, munro:kayla, toucan-hack, motherboard-teddy}. 

In this work, we analyze three commercially-available IoT toys, a hydration tracker, a smart pet, and a fitness band, for

\vspace{36pt}
\noindent application and network security and privacy vulnerabilities. 
All three products are clearly targeted towards minors on their respective websites and the Google Play store. 
The mobile application for the hydration tracker toy has 5,000+ installs from the Google Play store as of July 2018. The mobile applications for the smart pet and fitness band have 10,000+ and 1,000+ installs, respectively.
For comparison, the Hello Barbie companion app has 5,000+ installs.
All analyses in this paper were performed with the version of the toys and companion mobile applications available in November, 2017.

We chose these specific toys and mobile applications at the request of the state attorney general. 
We have not disclosed discovered vulnerabilities to the toy vendors due to ongoing attorney general investigations. 
The attorney general will notify the vendors when appropriate.
We keep the identities of the toys anonymous in this paper pending security patches.

We uncover several previously undisclosed vulnerabilities, including a lack of data encryption (HTTP instead of HTTPS), lack of authentication for accessing personally identifiable information (PII), POST token reuse, asymmetric HTTP responses allowing unique ID mining, PII in crash reports to third parties, and secret keys in source code constant files. 
We then identify how these vulnerabilities constitute violations of both COPPA and the individual toys' privacy policies. 

Despite their potentially serious impact, these vulnerabilities are all easily correctable. Their existence indicates substantial developer error, apathy, or ignorance of privacy and security best practices.
Even simple and well-known security practices, such as transport-layer encryption, are neglected by developers. 
These case studies motivate more rigorous auditing of Internet-connected toys by researchers and consumer advocacy groups as well as improved development practices by IoT toy manufacturers. 

This paper makes the following contributions:
\begin{itemize}
    \item Presents previously undisclosed vulnerabilities of three Internet-connected children's toys, informing general observations about the current state of smart toy security and privacy:
    \begin{itemize}
        \item Lack of industry-standard security practices, especially encryption/authentication of communications with first-party cloud services, leaves personal data unprotected and constitutes violations of manufacturer privacy policies and federal COPPA regulation.
        \item Use of common third-party analytics services across smart toys could allow cross-device tracking of child behavior.
    \end{itemize}
\end{itemize}
\vspace{12pt}

\section{Related Work}
\IEEEpubidadjcol
Recent high-profile hacks of smart toys \cite{guardian:barbie, motherboard-teddy}, have drawn considerable press attention to the privacy and security risks of Internet-connected devices for children \cite{wp-toys, wired-toys}.
The academic community has responded to consumer concern with studies auditing smart toys \cite{munro:kayla, valente2017security, jones2016can, pleban:drones}, analyzing children's and parents' interactions with smart toys \cite{mcreynolds2017toys, manches2015three}, and proposing technical and legal frameworks or requirements for improving smart toy security and privacy \cite{haynes2017framework, hung2016glance, yankson2017privacy, rafferty2017towards, yong:iotpedos}. 
This work fits into the general methodological framework of previous toy security audits in that it systematically explores vulnerabilities in consumer smart toys, provides proof-of-concept exploits, and suggests remediation techniques. However, unlike most previous studies, this work situates discovered vulnerabilities in the context of federal COPPA regulation and the examined toys' respective privacy policies. By doing so, it provides a glimpse into how IoT toy developers' practices can fail to match their own privacy promises or adhere to federal regulations. 
These results corroborate the findings of Reyes et al.~\cite{reyes2018won} that many Android applications targeted toward children are potentially in violation of COPPA.
\section{Method}
We evaluate the security of three commercially-available IoT toys: 
a hydration tracker, a smart pet, and a fitness band. 
None of the toys have publicly disclosed vulnerabilities.

We evaluate each toy with static and dynamic analyses to understand its  functionality and to discover application and network security vulnerabilities.  Finally, we check whether any discovered vulnerabilities violate the toys' privacy policies or COPPA.  

Our approach involves commonly used tools for penetration testing and security auditing, emphasizing the breadth and severity of smart toy vulnerabilities that can be discovered using these methods, which are publicly available and have a low barrier to use for potential attackers.
The vulnerabilities reported are unlikely comprehensive---future research using more sophisticated techniques, such as taint tracking~\cite{zhu2011tainteraser}, may uncover additional vulnerabilities; however, we focus on the highest impact issues that are both easily detectable by attackers and easily correctable by manufacturers.

\subsection{Static Analysis}
We collected publicly-available documentation about each toy, including user manuals and privacy policies. 
We examined the documentation to understand the capabilities of each toy and to learn which discovered vulnerabilities violate its privacy policy. 

We also decompiled the Android mobile application associated with each toy using the \textit{jadx} Dex to Java compiler \cite{jadx}. 
We examined the source code for security vulnerabilities, such as cleartext passwords, network serialization formats, and metadata about developers. 

\subsection{Dynamic Analysis}
\begin{figure}
\centering
\includegraphics[width=0.38\textwidth]{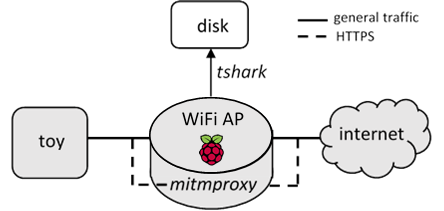}
\caption{Packet capture pipeline for recording IoT toy network traffic. All toys were connected to an instrumented WiFi access point implemented on a Raspberry Pi.}
\label{fig:pcappipeline}
\end{figure}

We collected packet captures from the toys by associating them with an instrumented WiFi access point implemented on a Raspberry Pi (Figure~\ref{fig:pcappipeline}) \cite{iot-inspector}. 
HTTPS traffic was decrypted using \textit{mitmproxy} \cite{mitmproxy}. 
The command line utility \textit{tshark} \cite{tshark} was run on the Raspberry Pi to log packets to disk. 
Recorded packet traces were then analyzed visually with \textit{wireshark} \cite{wireshark} and programmatically with the Python Scapy library~\cite{scapy}. 
We focused on traffic between the toys and backend cloud servers because it is easy to intercept and subject to the widest range of attacks. Security and privacy evaluations of non-WiFi local communications (Bluetooth, Zigbee, etc.) remain a topic for future work.

We also used the network mapper \textit{nmap} \cite{nmap} to determine what ports were open on the toys.
No port scans were launched against remote servers that the toys communicated with, as scanning permission was neither requested of nor granted from the respective site owners
\section{Hydration Tracker}
\label{hydrationtracker}
The hydration tracker consists of a physical water bottle designed for child use and an associated mobile application.
The hydration tracker was released in 2016, and the mobile application has 5,000+ installs from the Google Play store as of July 2018. An alternative version of the water bottle was released in 2018 with additional features, indicating continued sales of the toy.
The user, likely a child, can create a profile on the application, and after entering biographical and personal information such as name, age, weight, and height, a water consumption goal is computed. 
Water consumed from the bottle is logged and can be viewed from within the application. 
The water bottle has a graphical interface for monitoring daily consumption goals.
The user can also add a profile picture that appears inside the application. 
This photo is likely of the child, although privacy-conscious users may choose a non-identifying photo.

\subsection{Network Behavior}
The water bottle is equipped with an 802.11 b/g/n WiFi card, but no Bluetooth functionality.
The mobile application, by virtue of running on a smartphone, can communicate via WiFi and Bluetooth. 

Decompiling the mobile application source code revealed numerous imported libraries for communication with third-party analytics and performance monitoring services.
These services include Yahoo's Flurry Analytics\footnote{https://developer.yahoo.com/analytics/}, Google Analytics\footnote{https://analytics.google.com/}, Crashlytics\footnote{http://try.crashlytics.com}, and a Chinese analytics platform.\footnote{https://www.jiguang.cn} 

\begin{figure}
\centering
\includegraphics[width=0.49\textwidth]{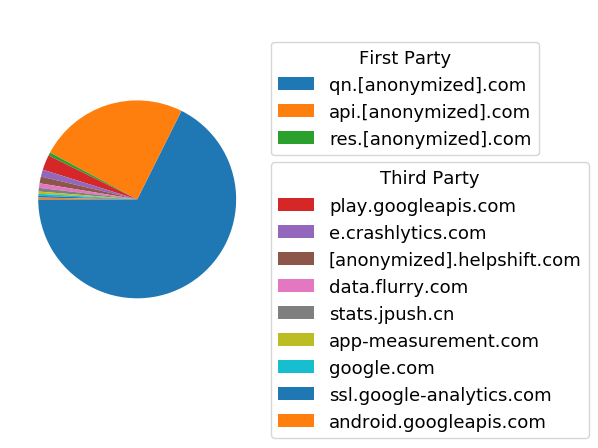}
\caption{Hosts communicating with the hydration tracker by percentage of bytes transferred over a 5-minute traffic recording.}
\label{fig:hydrationpichart}
\end{figure}

Summary flow statistics reveal that the hydration tracker communicates with approximately 12 remote hosts. Figure ~\ref{fig:hydrationpichart} plots these hosts weighted by how many bytes they exchanged with the hydration tracker.  
The hosts fall into two categories: servers belonging to the manufacturer of the toy and servers providing third-party analytics and performance monitoring.
All of the HTTP connections to third-party platforms were encrypted over SSL, and all connections to manufacturer-owned servers were unencrypted, vanilla HTTP. 

\subsection{Vulnerabilities}
\label{sec:hydration-vulns}
Analysis of the hydration tracker revealed several vulnerabilities, primarily due to lack of encryption and authentication of HTTP sessions.

\subsubsection{HTTP Communications with Cloud Servers}
All communications between the hydration tracker and manufacturer-owned servers occur via unencrypted and unauthenticated HTTP GET and POST requests. 
These communications include requests for static content, such as background images in the application, as well as for dynamic content related to user behaviors.
When the user drinks from the water bottle, the bottle makes a HTTP POST request to report the event. 
An attacker could easily observe the contents of these requests, learn when a user is interacting with the device, and/or spoof a response with arbitrary content (Figure ~\ref{fig:mal-behavior}a). 
This content could contain executable data that triggers remote code execution in the application or smart water bottle.

Unencrypted HTTP GET requests are also used to fetch the user's profile picture upon loading or restarting the mobile application. 
An adversary could eavesdrop on these communications and observe the profile photographs of children using the hydration tracker, a clear privacy breach (Figure ~\ref{fig:mal-behavior}b). 

\subsubsection{POST Token Reuse}
In addition to being unencrypted, POST requests from the hydration tracker reporting drinking events reuse an authentication token transmitted as an HTTP header (this appears to be a custom header as it is not specified in the relevant HTTP RFCs). An attacker could observe the token of a legitimate user by eavesdropping a valid request and then spoof valid requests to the server with arbitrary content (Figure ~\ref{fig:mal-behavior}c). This could allow remote code execution on the server and be used as an attack vector to dump the server's database filled with sensitive user data. 

Securing POST requests with HTTPS would not prevent request spoofing. If an attacker was able to guess a valid token, they could establish their own secure session to the server and continue spoofing requests pretending to be a legitimate user. The tokens used for these POST requests should have a time-to-live and be periodically refreshed to minimize the window of opportunity an attacker has to exploit a valid token.

\subsubsection{No Authentication for User Profile Photos}
HTTP GET requests for user profile pictures are not only unencrypted, but also unauthenticated. An attacker able to generate valid requests could receive the profile pictures of any hydration tracker user anywhere in the world, even if the attacker is not able to directly eavesdrop hydration tracker communications. 

\begin{figure}
\centering
\subfloat[A malicious actor observing an unencrypted GET request and spoofing the response with a malicious payload.]{
\includegraphics[width=0.35\textwidth]{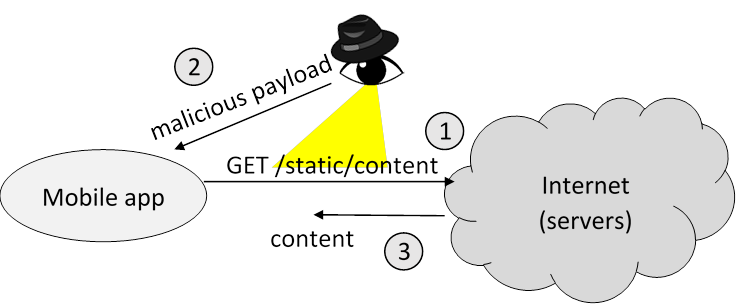}
}\\
\subfloat[A malicious actor observing personally-identifiable information in an unencrypted server response.]{
\includegraphics[width=0.35\textwidth]{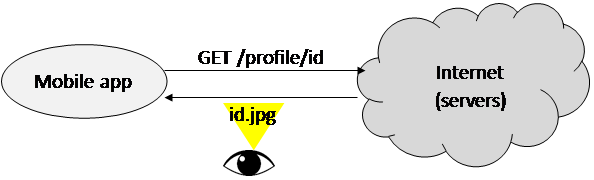}}\\
\subfloat[A malicious actor obtaining a valid token from a POST request and spoofing a new request to the server with a malicious payload.]{
\includegraphics[width=0.35\textwidth]{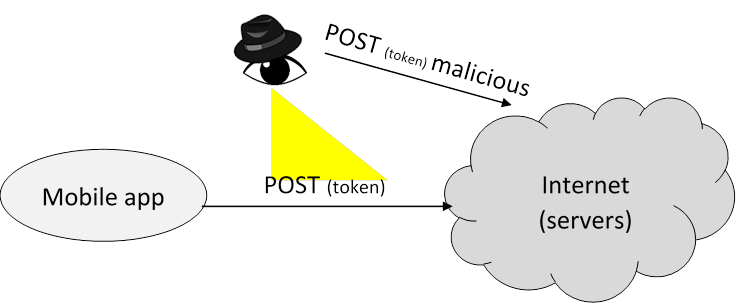}}\\
\subfloat[A malicious actor observing a GET request and spoofing it to the server to receive a response.]{
\includegraphics[width=0.35\textwidth]{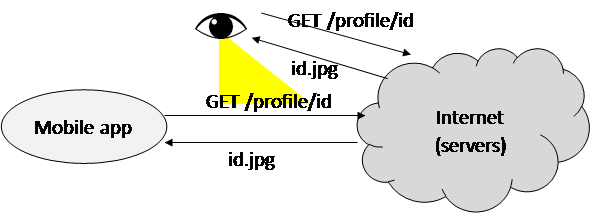}}
\caption{Malicious actions enabled by discovered security vulnerabilties.}
\label{fig:mal-behavior}
\end{figure}

Even if the hydration tracker were updated to use HTTPS, this vulnerability would not be solved. An attacker could just make an HTTPS connection to the server and spoof the GET request over an encrypted connection (Figure ~\ref{fig:mal-behavior}d). An authentication mechanism is necessary to verify that that user making the request has access rights to the profile picture. 

\subsubsection{Continued Availability of Overwritten Profile Photos}
After a user overwrites their current profile picture with a new one, their old profile picture is still accessible via the original HTTP GET request. This is functionally unnecessary, as there is currently no feature of the application to switch to a previous profile photo. Moreover, this continued availability makes the window of opportunity for an attacker to obtain a profile photo via sniffing or spoofing essentially infinite, in addition to needlessly polluting the space of valid photo tokens and thus increasing the chances of a random guess corresponding to a valid token. 

\subsubsection{Asymmetric HTTP Response Codes Allow Profile Photo ID Discovery}
HTTP GET requests for user profile picture contain a unique 12-character identification token. 
However, requests with a truncated 3-character token receive differentiated responses depending on whether the 3 characters are a prefix of a legitimate token.
Specifically, prefixes of real user tokens yield HTTP 301 Moved Permanently response codes,
while prefixes of invalid tokens yield HTTP 404 Not Found response codes. 
Combined with the previously described lack of authentication for profile photo GET requests, this implementation flaw substantially reduces the time required to guess valid picture tokens (Section ~\ref{exploit}).

\subsubsection{Personally Identifiable Information in Crash Reports}\label{hydrationmitmdisclosure}
\begin{figure}[t]
\centering
\includegraphics[width=0.49\textwidth]{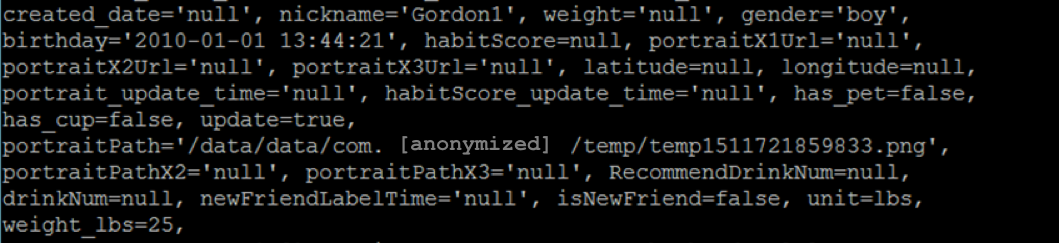}
\caption{Excerpt of crash report containing sensitive information sent to a third-party analytics service from the hydration tracker toy.}
\label{fig:hydrationcrash}
\end{figure}

During a transient network error, the mobile application sends a crash report to a third-party analytics platform containing personally identifiable information, including the name, gender, birthday, and weight of the user (Figure~\ref{fig:hydrationcrash}).
Although this information was sent encrypted over HTTPS, it is unlikely that a third-party analytics platform needs this sensitive data. 
The application should exclude these data from crash reports.

\subsection{Profile Picture Mining}\label{exploit}
This section outlines an attack against the hydration tracker to obtain profile pictures of arbitrary users without access to the device.

\begin{figure}
\centering
\includegraphics[width=0.49\textwidth]{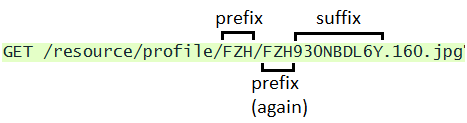}
\caption{Example HTTP GET request for hydration tracker user profile picture with 3-character prefix labeled. }
\label{fig:hydrationprofileget}
\end{figure}

When the hydration tracker mobile application launches, and periodically thereafter, it makes an unencrypted HTTP GET request containing a unique 12-character identification token to a manufacturer-owned server (Figure ~\ref{fig:hydrationprofileget}). If the token matches an existing user's picture, the server will respond with that picture. 

Because the request for the picture is unauthenticated, an attacker can simply send requests with tokens for pictures from arbitrary users. Although the attacker does not know which picture tokens correspond to existing photos, they can repeatedly attempt to guess valid tokens until the server returns a picture. 

A valid token consists of 12 characters, drawn from the uppercase alphanumeric alphabet.
The uppercase alphanumeric alphabet has $36$ entries, namely the $10$ digits and the $26$ uppercase letters A-Z. The number of valid tokens that can be created from this alphabet is $36^{12}$.

It is possible to discover whether 3-character strings are prefixes of existing tokens separately from discovering entire tokens.
The first 3 characters of a token are included twice in a GET request URL, first as a separate prefix string and then in the entire token (Figure~\ref{fig:hydrationprofileget}). 
An HTTP 301 response code indicates that a guessed prefix is valid, while an HTTP 404 indicates that the prefix is invalid. 
This ability to discover token prefixes considerably reduces the expected runtime to guess and verify tokens of existing user profile pictures. 

An attacker would need to test $36^3$ prefixes to discover all $p$ valid prefixes. This would take less than 2 hours for an HTTP response rate of 200ms.
The attacker would then need to test $p\times 36^9$ suffixes to obtain the profile photos of all users of the hydration tracker. 
This would take more than $10^5$ years with the same response rate---far too long for a reasonable attack. 
However, the attacker may be satisfied with finding only a fraction of all the profile pictures. In this case, the expected runtime of the algorithm scales down proportionally to the desired fraction.

Furthermore, an attacker could also parallelize the attack by testing guessed tokens on many machines. Care would have to be taken in seeding the pseudorandom number generators of different instances of the algorithm to minimize overlap during the permutation of the tokenspace. 

In spite of the combinatoric infeasibility of testing all possible picture tokens, the fact that any user profile photos could be recovered by an attacker
raises serious privacy concerns. 

\subsection{Violations of COPPA and Privacy Policy}
The hydration tracker violates COPPA in two ways. First and foremost, the profile picture mining attack and its enabling vulnerabilities allow unaffiliated actors to obtain profile pictures of arbitrary users without access to the device. Since this device is targeted towards children, and user profile photos are likely of children, this vulnerability is in clear violation of COPPA section 312.8, namely that the manufacturer must have "reasonable procedures to protect the confidentiality, security, and integrity of personal information collected from children" and "take reasonable steps to release children's personal information only to service providers and third parties who are capable of maintaining the confidentiality, security and integrity of such information"~\cite{coppa}.

The hydration tracker also violates COPPA's data retention policy. After changing the photo associated with a given profile to a new picture, the URL used to retrieve the old photo still works. This violates COPPA section 312.10, which mandates that the manufacturer "shall retain personal information collected online from a child for only as long as is reasonably necessary to fulfill the purpose for which the information was collected" and that the manufacturer "must delete such information using reasonable measures to protect against unauthorized access to, or use of, the information in connection with its deletion"~\cite{coppa}.

The hydration tracker also violates its own privacy policy with the vulnerabilities in Section~\ref{sec:hydration-vulns}. The privacy policy states that users' personal information is "contained behind secured networks\dots accessible by a limited number of persons who have special access rights" and is "encrypted via Secure Socket Layer (SSL) technology." These assertions are clearly false.
\section{Smart Pet}
The smart pet is a 
non-connected plush toy with a companion
mobile application, available on both iOS and Android platforms. Upon creating an account through the user interface, a virtual ``pet'' verbally prompts the user for their name and birthday. Afterwards, the user 
inserts the phone into the plush toy and
plays with it
in interactive skits. 
The smart pet was released in 2013, and the mobile application has 10,000+ installs from the Google Play store as of July 2018. The company which originally made the smart pet has since partnered with a major global toy manufacturer (top five worldwide by revenue) and continues active development of new smart pet versions.

\subsection{Network Behavior}
\begin{figure}
\centering
\includegraphics[width=0.49\textwidth]{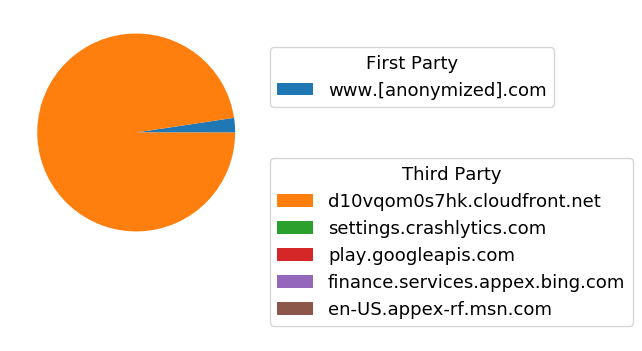}
\caption{Hosts communicating with smart pet by percentage of bytes transferred over a 4-minute traffic recording.}
\label{fig:smartpetpichart}
\end{figure}

Because the smart pet system is a standalone mobile application, it can speak the same network protocols as the smartphone. On the Android platform, the application requests permissions to access and change WiFi connectivity state. The description of the application in the Google Play store states that Bluetooth capability is necessary for certain features.
Decompiling the binary revealed two imported libraries for third party analytics platforms, Google Analytics and Crashlytics. 
Summary conversation and endpoint statistics reveal that the smart pet communicates with approximately 6 remote hosts (Figure ~\ref{fig:smartpetpichart}). 
The hosts are comprised of manufacturer-owned servers, third-party analytics platforms, Microsoft news services, and a large content distribution network. Sessions to Microsoft news domains (\texttt{msn.com}) are over unencrypted HTTP. Sessions to all other hosts are encrypted over SSL.

\subsection{Vulnerabilities}
The smart pet analysis revealed vulnerabilities involving constant storage, encryption, and authentication. 

\subsubsection{Constants File Exposes Sensitive Tokens}
\begin{figure}[t]
\centering
\includegraphics[width=0.49\textwidth]{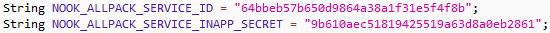}
\includegraphics[width=0.49\textwidth]{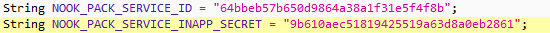}
\caption{Secret constants stored in plaintext in the smart pet source code.}
\label{fig:smartpetconstants}
\end{figure}

The source code of the smart pet contains a Java file with several constants in cleartext (Figure~\ref{fig:smartpetconstants}). Most of them are innocuous; however, two constants in particular appear to be secret tokens for the Nook in-app purchase API:\footnote{The Nook API is deprecated as of March 2016, however, this toy was available at least as early as Aug 2014, meaning this vulnerability was live for nearly 2 years.} 
\begin{itemize}
    \item NOOK\_ALLPACK\_SERVICE\_INAPP\_SECRET
    \item NOOK\_PACK\_SERVICE\_INAPP\_SECRET
\end{itemize} 
Rather than storing these tokens in plaintext in source code, 
these secret tokens should be stored in an encrypted format and only be decrypted in memory. 
Otherwise, an attacker could make API calls for in-app purchases on behalf of the client. 

\subsubsection{Cleartext HTTP for XML Files}
The mobile application makes unencrypted HTTP GET requests for XML files containing news headlines from several Microsoft domains. An attacker could intercept, read, and spoof these requests or responses with little effort. In the worst case, depending on how the XML is parsed and used, this could lead to remote code execution. Maliciously formatted XML has been known to cause remote code execution flaws in server software such as Apache Struts, a vulnerability which was
was exploited in the 2017 Equifax breach \cite{CVE-2017-5638}. 
The GET requests for these XML files should be made over HTTPS to mitigate response-spoofing attacks and snooping by malicious agents. 
\section{Fitness Band}
The fitness band consists of a wristband and an associated mobile application. 
After synchronizing the wristband to the mobile application, the user can play games inside the application. Characters in the game are controlled by movement of the arm wearing the wristband.
The fitness band was released in 2014, and the mobile application has 1,000+ installs from the Google Play store as of July 2018. Active development of the band seems to have stopped. The last update to the mobile application occurred in 2016, but the physical band is still available for purchase on Amazon.com as of July 2018.

\subsection{Network Behavior}
The wristband is Bluetooth enabled but has no WiFi capability.
The application, by virtue of running on a smartphone, can communicate via WiFi and Bluetooth. 

Decompiling the binary to source code revealed four imported third-party libraries for analytics and performance monitoring.
These include Yahoo's Flurry Analytics, Google Analytics, Crashlytics, and Unity 3D statistics.\footnote{https://hwstats.unity3d.com/}

\begin{figure}
\centering
\includegraphics[width=0.49\textwidth]{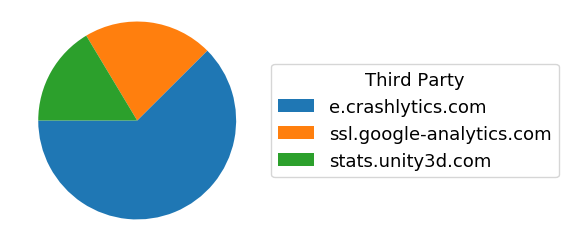}
\caption{Hosts communicating with fitness band by percentage of bytes transferred over a 4-minute traffic recording. No communications with first-party servers were observed.}
\label{fig:fitnesspichart}
\end{figure}

Summary conversation and endpoint statistics reveal that the fitness band communicates significantly with approximately 3 remote hosts (Figure ~\ref{fig:fitnesspichart}). 
The hosts are only third-party analytics platforms. All connections are encrypted using SSL.

\subsection{Vulnerabilities}
No vulnerabilities were discovered in the fitness band's decompiled source code or its small amount of communication with remote hosts.

\section{Discussion}
This section reflects on the current state of smart toy security based on the results of this study. 

\subsection{Privacy Policy Violations}
Smart toy manufacturers promise strong privacy and security guarantees explicitly in their own privacy policies and implicitly by adhering to COPPA, but fall short during implementation. For example, the hydration tracker's privacy policy claims that personally-identifiable information is protected "behind secured networks\dots accessible by a limited number of persons who have special access rights" and "encrypted via Secure Socket Layer (SSL) technology."
As shown in Section~\ref{hydrationtracker}, this is clearly not the case. Furthermore, this pattern of good privacy posture being let down by weak implementation is not unique to the toys in our study. Recently, researchers have shown that in spite of an encryption mechanism intended to preserve the privacy of children's voice recordings in transit, a particular toy's usage of a fixed set of keys made it possible to completely decrypt any voice communication between the toy and its server \cite{valente2017security}.

Such violations invite legal action against the toy manufacturers. In January 2018, the FTC settled charges against toy manufacturer VTech for ``failing to take reasonable steps to secure the [children's] data it collected" and other COPPA violations \cite{coppa-vtech}. In addition to federal action, state attorneys general have settled with technology companies for COPPA and privacy policy violations \cite{nys-truste}.  Our results should encourage researchers to continue performing security analyses of Internet-connected products and to provide information about discovered vulnerabilities to the FTC or state attorneys general. 

\subsection{Overlapping Third-Party Analytics}
Each smart toy talks to a similar set of third-party analytics and performance monitoring platforms. Google Analytics and Crashlytics libraries were included in the source code of all three toys. Yahoo's Flurry Analytics was included in the source code of two of the toys. This suggests that a small set of platforms have high visibility into a broad set of smart toys. Coupled with over-reporting of personally-identifiable information to analytics services (Section ~\ref{hydrationmitmdisclosure}), these platforms could be receiving and storing more sensitive data than users expect.

This privacy vulnerability is similar to well-studied concerns about third-party web tracking. The incorporation of third-party content, such as advertisements or social media integration, on numerous web sites allows third-party companies to track individuals' browsing behavior across large portions of the web \cite{mayer2012third}. 
If third-party analytics companies receive data from numerous smart toys, they could similarly construct detailed profiles of individual children's behaviors, as well as large-scale datasets about smart toy users of value to manufacturers and distributors. Further research should investigate the proliferation of these analytics platforms across a wider set of smart toys, as well as the extent of user information they collect. 

\subsection{Lack of Encryption with First-Party Servers}
We found several network security vulnerabilities in communications between smart toy systems and first-party manufacturer-owned servers, most due to lack of encryption or authentication. In comparison, communications between smart toy systems and third-party analytics servers were all secure. This is likely because third-party analytics libraries tend to be from mature companies that use encryption and authentication best practices, while first-party communications are implemented by toy manufacturers without the time, resources, or know-how to use encrypted protocols. This is a phenomenon seen across the IoT space, as manufacturers of physical objects with little software development experience begin creating ``smart'' products. 

This indicates an analogy to the early days of the web, where encryption and other security best practices were much less well-known and widely adopted. Web security has gradually improved due to a variety of factors, including site hosting services with security baked-in, improved web development libraries reducing the potential for programming errors and simplifying SSL/TLS support, improved browser-based security features, new versions of web standards, and more. This comparison invites IoT researchers and manufacturers to look to the systems that have improved web security as inspiration for simplifying and securing IoT development.

This observation also supports the role that regulation can play in securing smart toys and other IoT devices.  The European Union's General Data Privacy Regulation (GDPR) encourages, but does not mandate, the use of encryption to protect infringement (Recital 83)~\cite{venafi-gdpr}.
Industry advisory groups have made similar recommendations for IoT device design and development, including using ``strong authentication by default'' and following ``security and cryptography best practices''~\cite{bitag}. Mandating SSL/TLS or such best practices would prevent vendors with little security background from releasing devices with egregious cleartext vulnerabilities.  The availability of SSL/TLS libraries in every commonly used language for IoT development would already prevent such regulations from being overly burdensome to manufacturers. Excessive manufacturer effort to implement security features can no longer be an excuse for insecure Internet-connected toys that place children at risk.
\section{Future Directions}
While this study discovered several security vulnerabilities and privacy policy violations, additional research is needed to protect children from insecure Internet-connected toys.

\subsection{Analyzing Additional Toys}
The three toys we analyzed demonstrate the breadth of network security vulnerabilities present in Internet-connected children's products and detectable using standard analysis techniques. While these case studies provide further evidence that IoT manufacturers need to place higher priority on security and indicate specific problem areas for attention, a comprehensive view of smart toy security would require analyzing many additional toys and repeating these analyses as the toys and corresponding mobile applications receive software updates.

Since this project was focused on the three toys requested by the state attorney general, we leave repeated analyses of additional toys for future work. The development of an automated auditing tool, as discussed in the following section, would also greatly simplify the process of analyzing many toys in parallel.

\subsection{Automated Toy Auditing}
The static and dynamic analyses we performed in this paper mostly involved manual or custom-programmed data inspection, whether the data was decompiled source code, privacy policies, or network traffic traces. 
Researchers have long proposed and developed automated network security analysis tools, both for IoT devices~\cite{simpson2017securing} and more traditional network configurations~\cite{roesch1999snort, ou2005mulval}.
There are also many commercially-available analysis tools which perform online and offline security audits of network traffic or source code.
However, none of these tools connect network or code analysis with product privacy policies. The development of an automated analysis tool for identifying device privacy and security vulnerabilities and determining whether they constitute privacy policy breaches would greatly reduce regulator burden, especially in the IoT market where new devices and device updates are released faster than they can be manually audited. However, creating such a tool would be non-trivial, because privacy policies are often written with intentionally vague wording that does not directly map to concrete device behaviors~\cite{bhatia2016theory}.
A successful automated tool would need to map privacy policy text to verifiable security or privacy properties and check devices for these properties.

\section{Conclusion}
This work has examined three commercially-available IoT toys to gain a deeper understanding of the smart toy security and privacy landscape. Through a combination of static and dynamic analysis, several previously undisclosed vulnerabilities were discovered, including neglected encryption and authentication, POST token reuse, sensitive user information in crash reports, and secret keys in source code. 
These vulnerabilities violate the toys' individual privacy policies as well as federal COPPA regulations for handling children's data.
Additionally, a small set of third-party analytics platforms receives data from all examined toys, possibly allowing detailed user data collection similar to third-party web tracking. 
These results indicate that Internet-connected children's toys require continued security and privacy auditing, and that further work is needed to help IoT toy manufacturers improve security and privacy development practices.  

\section*{Acknowledgment}
This work was supported by a Google Faculty Research Award, the National Science Foundation through awards CNS-1535796 and CNS-1539902, and the Princeton University Center for Information Technology Policy Internet of Things Consortium.

\IEEEtriggeratref{18}
\bibliographystyle{IEEEtran}
\bibliography{IEEEabrv,IoT-Toys}

\end{document}